\documentclass{tMOP2e}

\usepackage{epstopdf}
\usepackage{subfigure}
\usepackage{dsfont}
\theoremstyle{plain}

\theoremstyle{definition}

\theoremstyle{remark}

\newcommand{\ket}[2][]{{|#2\rangle_{#1}}}
\newcommand{\bra}[2][]{{}_{#1}\langle #2|}
\newcommand{\braket}[3][]{{{}_{#1}\langle#2|#3\rangle_{#1}}}
\newcommand{\proj}[2][]{\ket{#2}_{#1}\bra{#2}}

\begin{document}



\title{Quantum memory receiver for superadditive communication\\using binary coherent states}

\author{
\name{Aleksandra Klimek, Micha{\l} Jachura, Wojciech Wasilewski, and Konrad Banaszek$^{\ast}$\thanks{$^\ast$Corresponding author. Email: Konrad.Banaszek@fuw.edu.pl}}
\affil{Wydzia{\l} Fizyki, Uniwersytet Warszawski, ul.\ Pasteura 5, 02-093 Warszawa, Poland}
\received{v5.0 released January 2015}
}

\maketitle

\begin{abstract}
We propose a simple architecture based on multimode quantum memories for collective readout of classical information keyed using a pair coherent states, exemplified by the well-known binary phase shift keying format. Such a configuration enables demonstration of the superadditivity effect in classical communication over quantum channels, where the transmission rate becomes enhanced through joint detection applied to multiple channel uses. The proposed scheme relies on the recently introduced idea to prepare Hadamard sequences of input symbols that are mapped by a linear optical transformation onto the pulse position modulation format [Guha, S. {\em  Phys. Rev. Lett.}\ {\bf 2011}, {\em 106}, 240502]. We analyze two versions of readout based on direct detection and an optional Dolinar receiver which implements the minimum-error measurement for individual detection of a binary coherent state alphabet.
\end{abstract}


\section{Introduction}
One of the striking consequences of the quantum nature of physical systems is the impossibility to discriminate perfectly their states that are non-orthogonal in terms of the scalar product between the corresponding state vectors \cite{Chefles2000}. This fact has profound implications for secret communication in the form of quantum key distribution protocols \cite{QKD}, but it also leads to non-trivial effects when transmission of classical information is considered \cite{Gallion2009}. In optical communication, standard schemes to encode a stream of bits employ a pair of coherent states, e.g. the vacuum state and a coherent state with a non-zero amplitude in the case of on-off keying (OOK), or two coherent states with equal amplitudes but opposite phases in binary phase shift keying (BPSK) \cite{Kitayama2014}. When an energy constraint is imposed in the above schemes, the error rate grows with the decreasing signal power, as the two coherent states encoding the bit value become less and less distinguishable in the quantum mechanical sense. An intriguing strategy to boost throughput in such a case is to employ collective detection of the received signal, which for very weak signals can even qualitatively enhance the scaling of the attainable transmission rate with the mean power. The fundamental reason behind this enhancement is that a quantum measurement provides in general only partial knowledge about the state of the measured system and collective detection of multiple elementary systems can be designed to reveal more relevant information \cite{Sasaki1998,BanaszekNPH2012}.

An elegant scheme to achieve superadditivity for binary phase shift keyed signals has been recently described by Guha \cite{Guha2011}. The basic idea is to prepare sequences of BPSK symbols that can be mapped using a linear optical circuit onto the pulse position modulation (PPM) format. This format can be read out using direct detection. Moreover, with the right choice of the sequence length \cite{ChungGuha2013,WangWornell2014,KochWang2014,Jarzyna2015} this strategy approaches in the leading order the capacity of a narrowband bosonic channel for low signal powers \cite{GiovGarcNPH2014}. The purpose of the present contribution is to propose an implementation of the linear circuit processing BPSK sequences in a multimode quantum memory interface \cite{Nunn2008,Chrapkiewicz2012,Dabrowski2014}.
The proposal is motivated by recent demonstrations of fully controllable linear transformations between atomic spin coherences and optical fields \cite{Campbell2012,Campbell2014}.
This approach would be well suited to process sequences transmitted in a single spatial mode and encompassing multiple time bins. The presented scheme points to possible applications of quantum memories not only in quantum information processing, but also in future optical communication systems operated at the quantum limit.

This paper is organized as follows. First, in Sec.~\ref{Sec:BPSK} we review the principle of BPSK and the attainable transmission rates in the low power regime. The strategy to achieve superadditivity using sequences of BPSK symbols is summarized in Sec.~\ref{Sec:Hadamard}. The proposal for the quantum memory interface to process BPSK sequences is described in Sec.~\ref{Sec:QM}. For short sequence lengths, we analyze in Sec.~\ref{Sec:Hybrid} possible gains from the application of a minimum-error Dolinar receiver at one of the output ports. Finally, Sec.~\ref{Sec:Conclusions} concludes the paper.

\section{Binary phase shift keying}
\label{Sec:BPSK}

Any two coherent states can be mapped via a unitary linear optical transformation onto a pair with the same mean photon number but opposite phases.
This transformation can be realized using a beam splitter with transmission approaching one and an auxiliary coherent field \cite{BanaWodkPRL1996}.
Moreover, if both the states are equiprobable such a pair minimizes the mean energy for a fixed separation between the complex amplitudes of the coherent states, characterizing their distinguishability. Therefore in the following we will restrict our attention to this special case, commonly known in optical communication under the acronym BPSK. In simple terms, bits are encoded in the sign of the complex amplitude $\pm\alpha$ of coherent pulses, prepared with the same
mean photon number $\bar{n}= |\alpha|^2$ in each use of the channel. For large mean photon numbers $\bar{n}$, the two coherent states $\ket{\alpha}$ and $\ket{-\alpha}$ describing the pulses are almost orthogonal and the bit value can be read out with a negligible error using e.g.\ homodyne detection. Readout becomes less trivial in the regime of low mean photon numbers, when $\bar{n} \ll 1$, as the quantum mechanical scalar product between the two coherent states used for communication is then substantially nonzero, $|\braket{\alpha}{-\alpha}| = e^{-2\bar{n}}$, and therefore they cannot be distinguished with certainty \cite{Wittmann2008}.

The usefulness of a communication scheme for classical information transmission can be characterized with mutual information, which describes the strength of correlations between system preparations at the channel input and measurement results at the channel output. Importantly, mutual information provides the upper limit on the attainable transmission rate for a given communication scheme \cite{CoverThomas}. When two equiprobable quantum states are used as preparations and the physical systems transmitted in consecutive channel uses are measured individually, the optimal detection strategy is to apply the minimum-error measurement described  by Helstrom \cite{Helstrom1976,TomassoniParis}. From the information theoretic point of view, such a scheme is described by a binary symmetric channel with the error rate given by
\begin{equation}
\label{Eq:ErrorHelstrom}
\varepsilon(\bar{n}) = \frac{1}{2}(1- \sqrt{1- |\braket{\alpha}{-\alpha}|^2 }) = \frac{1}{2}(1- \sqrt{1- e^{-4\bar{n}} }).
\end{equation}
In the above expression we explicitly used the two coherent states constituting the BPSK alphabet. For a binary symmetric channel representing individual detection mutual information reads:
\begin{equation}
\label{Eq:IHel}
{\sf I}_{\text{ind}} = 1 - {\sf H} \bigl(\varepsilon(\bar{n}) \bigr) \approx \beta \bar{n},
\end{equation}
where ${\sf H}(x) = - x \log_2 x - (1-x) \log_2 (1-x)$ is the binary entropy measured in bits. The second approximate expression in Eq.~(\ref{Eq:IHel}) results from expanding mutual information up to the linear term in $\bar{n}$ and is valid in the regime $\bar{n} \ll 1$, with the proportionality constant equal to $\beta = 2/\ln 2 \approx 2.885$. In the case of two coherent states, the minimum-error measurement has a feasible implementation of the form of the Dolinar receiver \cite{Dolinar1976,Cook2007} comprising an auxiliary coherent reference beam, linear optics, photon counting and fast feedback loop to modulate the auxiliary beam.

For general measurement strategies on received systems, including collective detection, an upper bound on mutual information is given by the Holevo quantity $\chi$, which is defined mathematically as the difference between the von Neumann entropy ${\sf S}(\cdot)$ of the average state emerging from the channel and the average entropy of individual output states \cite{Holevo}. In the case of BPSK modulation, because individual states remain pure after transmission, the Holevo quantity is equal to the entropy of the statistical mixture of the two coherent states:
\begin{equation}
\label{Eq:HolevoBPSK}
{\chi} = {\sf S} \bigl( {\textstyle \frac{1}{2}} \proj{\alpha} + {\textstyle \frac{1}{2}} \proj{-\alpha} \bigr)
= {\sf H} \bigl( {\textstyle\frac{1}{2}} (1- |\braket{\alpha}{-\alpha}|)\bigr) \approx \bar{n} \log_2 \frac{1}{\bar{n}}.
\end{equation}
The last expression, specifying the leading term in the expansion when $\bar{n} \ll 1$, shows that compared to individual measurements, collective detection enables a qualitative change in the scaling of attainable information with $\bar{n}$. Furthermore, the Holevo quantity calculated in Eq.~(\ref{Eq:HolevoBPSK}) approaches asymptotically for $\bar{n} \rightarrow 0$ in the leading order the classical capacity of a single-mode bosonic channel \cite{GiovGarcNPH2014}. These results point to substantial benefits of collective detection in the regime of low mean photon number.

\section{Hadamard sequences}
\label{Sec:Hadamard}

\begin{figure}
\begin{center}
\resizebox*{14cm}{!}{\includegraphics{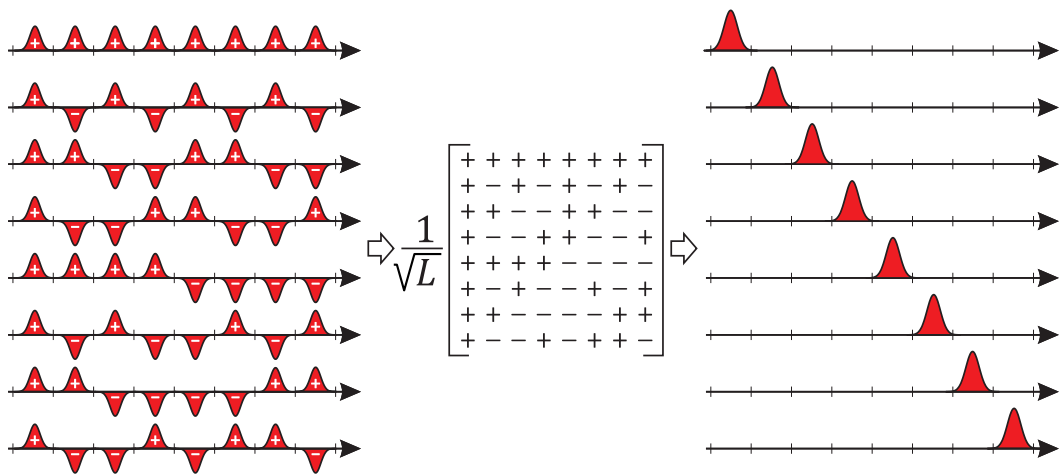}}

\vspace{5mm}

\begin{minipage}{110mm}
\caption{An exemplary superadditive communication scheme using the BPSK format for the sequence length $L=8$. The sender prepares sequences of BPSK symbols with $\pm$ signs defined by rows of a Hadamard matrix. At the receiver side, the symbols are interfered using a linear circuit described by a Hadamard matrix rescaled by $1/\sqrt{L}$. This maps the BPSK sequences onto the pulse position format where only one bin contains a pulse carrying the energy of the entire sequence. The position of the pulse
identifies unambiguously the received sequence.} \label{Fig:BPSK}
\end{minipage}
\end{center}
\end{figure}

Although general strategies to construct collective measurements approaching the Holevo quantity have been given theoretically \cite{Lloyd2011}, the challenge is to design joint detection schemes that could be implemented in practice using viable components. For BPSK modulation, a very elegant scalable scheme for sequence lengths $L$ equal to integer powers of two has been described by Guha \cite{Guha2011}. The basic idea, shown schematically in Fig.~\ref{Fig:BPSK}, is to select from all $2^L$ combinations of BPSK symbols only $L$ sequences that correspond to rows of a Hadamard matrix of dimension $L$. Hadamard matrices are symmetric with binary entries $\pm 1$, and their rows (or equivalently columns) form mutually orthogonal vectors \cite{Hedayat}. Collective detection of such {\em Hadamard words} is facilitated by an observation that rescaling a Hadamard matrix by $1/\sqrt{L}$ yields an orthogonal matrix which can be in principle implemented as a linear optical circuit. Because of the orthogonality property, each Hadamard word fed into the circuit will generate a non-zero pulse only in one output port of the circuit, different for each sequence, while all other ports will remain dark.

The above scheme effectively converts Hadamard BPSK words into the well known pulse position modulation (PPM) format, in which information is encoded in the position of a single pulse in the total number of $L$ otherwise empty bins. The most obvious strategy to read out the position of the pulse is to employ direct detection. Assuming ideal, unit-efficiency photon counting detectors without dark counts, either the position of the pulse is identified unambiguously, or this information is erased if no count is generated for any bin. From the information theoretic perspective such a communication scheme corresponds to the well-known erasure channel \cite{CoverThomas}, for which mutual information per one bin reads:
\begin{equation}
\label{Eq:IPPM}
{\sf I}_{\text{PPM}} = \frac{p}{L} \log_2 L.
\end{equation}
where $p$ is the probability of detecting the position of the pulse. In our case, because all $L$ BPSK states interfere constructively at one output port of the Hadamard circuit producing a pulse with the mean photon number $L\bar{n}$, the probability $p$ is given by
\begin{equation}
\label{Eq:pclick}
p = 1 - e^{-L\bar{n}}.
\end{equation}
Expanding the above expression up to the first order yields $p \approx L\bar{n}$, which implies that
\begin{equation}
\label{Eq:IPPMapprox}
{\sf I}_{\text{PPM}} \approx \bar{n} \log_2 L.
\end{equation}
This value is higher than the Helstrom limit for individual detection ${\sf I}_{\text{ind}}  \approx \beta \bar{n}$
when $L > 2^{\beta} \approx 7.4$. Therefore in the case of very weak pulses superadditivity is obtained for the minimum sequence length $L=8$.

It is worth to emphasize that the simple formula in Eq.~(\ref{Eq:IPPMapprox}) is valid only for $L\bar{n} \ll 1$, as for larger mean photon numbers the probability $p$ saturates at one. The exact expression given in Eq.~(\ref{Eq:IPPM}) has a well defined maximum as a function of $L$, which can be approximately identified by expanding $p$ up to the second order in $L\bar{n}$.
Mutual information ${\sf I}_{\text{PPM}}$ evaluated at this maximum has the expansion in $\bar{n}\ll 1$ of the form
\cite{ChungGuha2013,WangWornell2014,KochWang2014,Jarzyna2015}
\begin{equation}
{\sf I}_{\text{PPM}} \approx \bar{n} \log_2\frac{1}{\bar{n}} - \bar{n} \log_2 \ln \frac{1}{\bar{n}}.
\end{equation}
On the other hand, the capacity of a narrowband bosonic channel is given up to the second order as $\bar{n} \log_2 (1/\bar{n}) + \bar{n}/\ln 2$ for low signal powers and it coincides with the Holevo quantity for BPSK modulation found in Eq.~(\ref{Eq:HolevoBPSK}). It is seen that although the leading orders of both expressions are the same, the first order corrections exhibit different behaviour.

\section{Quantum memory implementation}
\label{Sec:QM}

In many commonly used optical communication links, e.g.\ fibres operating at telecom wavelengths, pulse sequences are transmitted in a single spatial mode. In this case, collective measurements described in the preceding section need to be implemented over multiple time bins. This requires synchronization of individual incoming pulses at the receiver while retaining mutual phase relations. One possible solution would be to employ fast optical switches and delay lines to equalize pulse arrival times before the Hadamard circuit. An alternative is to use quantum memories to transform coherently \cite{Reim2012} the incoming pulses into the PPM format. Within this approach the Hadamard circuit can be implemented piecewise with the incoming pulses using beamsplitter-type operations between light pulses and quantum memory modes \cite{Campbell2012,Campbell2014,Parniak2015}.

\begin{figure}
\begin{center}
\resizebox*{12cm}{!}{\includegraphics{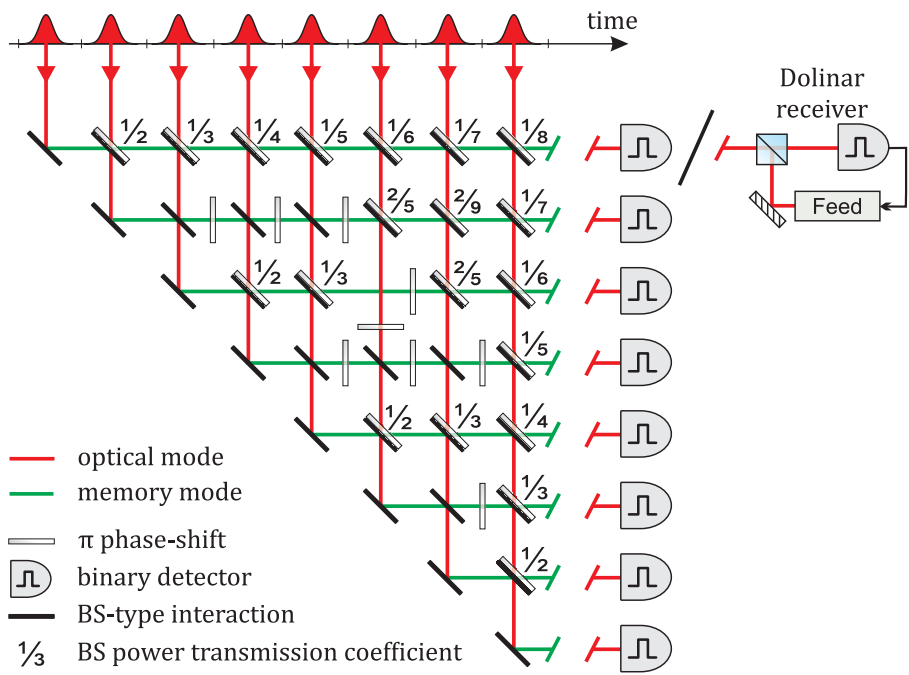}}

\vspace{5mm}

\begin{minipage}{110mm}
\caption{A quantum memory interface for converting Hadamard sequences of BPSK symbols into the PPM format shown schematically for $L=8$ sequence length. The horizontal axis represents the time flow. Arriving pulses interact with initially unoccupied memory modes depicted as horizontal lines. Black diagonal bars indicate beamsplitter-type interactions with the $\pi$ phase shift introduced for transmissions in both directions and reflections from shaded sides. Horizontal and vertical bars are additional $\pi$ phase shifts. Fractions labelling bars indicate power reflection coefficients. Unlabeled diagonal bars correspond to perfect reflections. At the output the memory modes are read out using direct detection. The case when one of the detectors is replaced by the Dolinar receiver is discussed in Sec.~\ref{Sec:Hybrid}.
} \label{fig:TriangleForm}
\end{minipage}
\end{center}
\end{figure}

A natural decomposition of the Hadamard circuit in this implementation is the triangular form of a general linear optical transformation discussed by Reck {\em et al.} \cite{Reck1994}. Its explicit form is shown schematically in Fig.~\ref{fig:TriangleForm} in the case of $L=8$ time bins. The first pulse is mapped onto a quantum memory mode. The $l$th pulse, $l=2,\ldots, L$, goes through $l$ quantum memory modes, as symbolized
by vertical lines in Fig.~\ref{fig:TriangleForm}. In each memory mode a transformation combining
the incoming light with the already stored excitation is driven by
suitable control fields \cite{Campbell2012}. The final $L$th memory, empty so far, is driven so as to store all
incoming light. At the end, the $L$ quantum memory modes contain the received sequence converted into the PPM format.
Detection of the excitations stored in the memories can be achieved
for example by mapping their contents back onto light and counting
optical photons in the standard manner.

\section{Hybrid detection}
\label{Sec:Hybrid}

Using Hadamard words constructed from BPSK symbols as described in Sec.~\ref{Sec:Hadamard}, superadditivity in mutual information can be demonstrated for at least $L=8$ time bins. On the other hand, a very simple hybrid scheme has been proposed for $L=2$ bins, where two consecutive pulses are interfered on a $50/50$ beam splitter with output ports monitored by a Dolinar receiver and a photon counting detector \cite{Guha2011}. With the right choice of probabilities of input sequences, the relative enhancement in mutual information is $2.5\%$, which is close to the value $2.8\%$ found numerically by optimizing joint two-system measurements \cite{Buck2000}. We will now discuss generalization of the hybrid scheme to more than two pulses.

The basic idea is to supplement the set of Hadamard words by a sequence $--\ldots-$. This sequence provides a non-zero pulse at the same output port of the Hadamard circuit as $++\ldots+$, but with the opposite phase. We will assume that this port is monitored by a Dolinar receiver, and both sequences are prepared with equal probabilities $(1-\lambda)/2$, where $0\le \lambda \le 1$. The remaining Hadamard words are sent with identical probabilities $\lambda/(L-1)$. The Hadamard circuit directs them to other output ports, each monitored with a photon counting detector. A click on a photon counting detector unambiguously identifies the input Hadamard word. If none of the photon counting detectors clicks, information from the Dolinar receiver is used. In this case, sequences $++\ldots+$ and $--\ldots-$ are identified with an error $\varepsilon(L\bar{n})$, because the mean total photon number in the entire sequence is $L\bar{n}$. Any other Hadamard word generates either measurement result on the Dolinar receiver with the same probability $1/2$, i.e.\ no information is obtained.

Mutual information for the above communication scheme can be cast in the following form:
\begin{equation}
\label{Eq:IHybrid}
{\sf I} = \frac{1}{L} \{ (1-\lambda) [1-{\sf H}( \varepsilon(L\bar{n}) )] + \lambda p \log_2 (L-1) + {\sf H}(\lambda p) - \lambda {\sf H}(p)\}.
\end{equation}
The overall multiplicative factor $1/L$ stems from rescaling mutual information per one time bin.
Within curly brackets, three contributions can be identified. The first one, given by $1-{\sf H}( \varepsilon(L\bar{n}))$ is mutual information for a binary symmetric channel with the error rate $\varepsilon(L\bar{n})$ corresponding to a minimum-error measurement on sequences $++\ldots+$ and $--\ldots-$. This contribution enters with the weight $1-\lambda$, which is the overall probability of preparing either sequence. The second term, $p\log_2(L-1)$, specifies mutual information for an $(L-1)$-ary erasure channel with the non-erasure probability $p$. This channel describes situation when any other Hadamard sequence is used, which occurs with the overall probability $\lambda$. Finally, the combination of the last two terms, ${\sf H}(\lambda p) - \lambda {\sf H}(p)$, specifies mutual information for the so-called $Z$ channel with a binary set of input symbols, when one symbol used with probability $\lambda$ is identified correctly with the probability $p$, whereas in the remaining $1-p$ fraction of cases it gives the same result as the second symbol, used with the probability $1-\lambda$. In optical communication, such a channel describes on-off keying where either a pulse or an empty bin are sent in each channel use, and an ideal photon counting detector without dark counts is used at the output.

\begin{figure}
\begin{center}
\resizebox*{11cm}{!}{\includegraphics{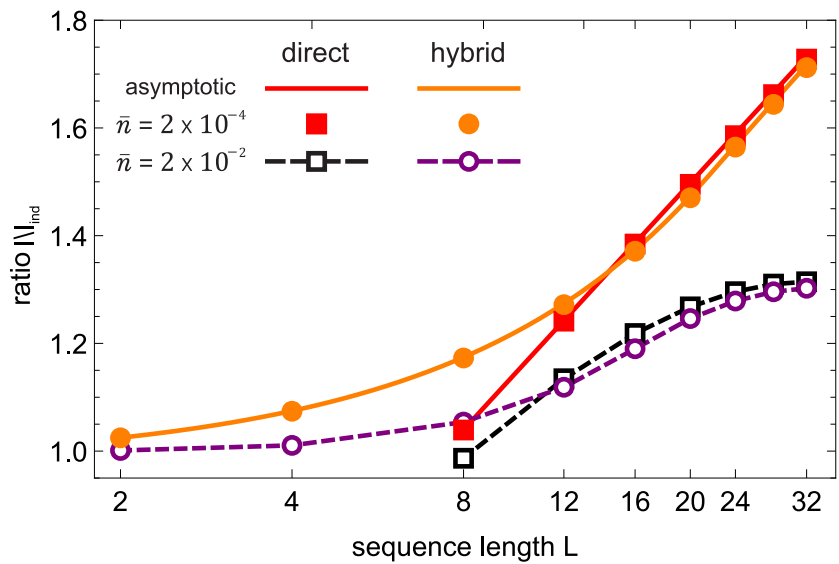}}

\vspace{5mm}

\begin{minipage}{110mm}
\caption{The ratio ${\sf I}/{\sf I}_{\text{ind}}$ of mutual information per bin for collective detection compared to the optimal individual detection case evaluated in Eq.~(\ref{Eq:IHel}).
Solid lines depict asymptotic results given in Eq.~(\ref{Eq:IPPMapprox}) for direct detection (gray solid line, red online) and in Eq.~(\ref{Eq:IHybridOptimal}) for hybrid detection (light gray solid line, orange online), with $L$ treated as a continuous parameter. Numerical results based on the exact expressions for the error probability in Eq.~(\ref{Eq:ErrorHelstrom}) and the count probability in Eq.~(\ref{Eq:pclick}) are shown for $\bar{n} = 2 \times 10^{-4}$ (filled symbols) and $\bar{n} = 2 \times 10^{-2}$ (empty symbols) in the case of direct detection (squares) and hybrid detection (circles). The dashed lines serve as guides to the eye. All sequence lengths $L\le 32$ for which Hadamard matrices exist have been included in the calculations.} \label{Fig:Hybrid}
\end{minipage}
\end{center}
\end{figure}

In our case the probability $p$ of a detector click is given by Eq.~(\ref{Eq:pclick}). Assuming that $p\ll 1$ we can approximate
\begin{equation}
{\sf H}(\lambda p) - \lambda {\sf H}(p) \approx \lambda p \log_2 \frac{1}{\lambda}.
\end{equation}
It is worth noting that the formula on the right hand side is formally equivalent to mutual information for the pulse position modulation format with $1/\lambda$ input words specified in Eq.~(\ref{Eq:IPPM}). In order to simplify calculations, in Eq.~(\ref{Eq:IHybrid}) we will expand up to linear terms in $\bar{n}$ the expressions for $p \approx L\bar{n}$ and $1-{\sf H}( \varepsilon(L\bar{n}) ) \approx \beta L\bar{n}$. After applying these approximations it is easy to find the optimal value of $\lambda$, which taking into account the constraint $0\le \lambda \le 1$ gives the following asymptotic expression for mutual information in the case of hybrid detection when $\bar{n} \ll 1$:
\begin{equation}
\label{Eq:IHybridOptimal}
{\sf I} = \begin{cases} {\displaystyle \bar{n} \left( \beta + \frac{L-1}{e 2^{\beta} \ln 2} \right),}  & \mbox{if $L < e 2^\beta +1 $} \\
\bar{n} \log_2 (L-1), & \mbox{if $L \ge e 2^\beta +1 $} \end{cases}
\end{equation}
It is seen that for large $L$ we recover the expression for $(L-1)$-ary pulse position modulation, as then the optimal strategy is not to use words $++\ldots+$ and $--\ldots-$ at all. In these cases direct detection scheme yields higher mutual information. However, enhancement is possible for short sequence lengths, as shown in Fig.~\ref{Fig:Hybrid} depicting the ratio ${\sf I}/{\sf I}_{\text{ind}}$. In the plots, we used two values of the mean photon number: $\bar{n}=2 \times 10^{-4}$ when the asymptotic expression given in Eq.~(\ref{Eq:IHybridOptimal}) is hardly distinguishable within the resolution of the graph from numerical results, and $\bar{n}=2 \times 10^{-2}$, which allows us to identify deviations from the asymptotics with the increasing mean photon number. It is seen that for larger $\bar{n}$ the superadditivity effect diminishes. In the case of direct detection one can notice that  mutual information ${\sf I}_{\text{PPM}}$ approaches a maximum with the increasing sequence length $L$, which is simply a result of the saturation of the count probability $p$ defined in Eq.~(\ref{Eq:pclick}).

\section{Conclusions}
\label{Sec:Conclusions}

We described theoretically a construction of a collective receiver for BPSK signal based on beam-splitter type transformations between incoming light pulses and quantum memory modes. Such a receiver can be used to demonstrate the superadditivity effect in classical communication over a quantum channel, with enhancement most strongly pronounced in the low-power limit. An interesting extension of the presented work may be to go beyond a sequence of time bins and to consider mixed time-frequency encodings within the available spectral bandwidth which could also be handled by architectures based on quantum memories \cite{Humphreys2014}.

Multimode interference underlying the superadditivity of the presented receiver relies on perfect phase and amplitude matching between interfering pulses. A recent study suggests that the collective BPSK detection scheme based on Hadamard words may be robust against moderate levels of phase noise \cite{JarzynaKlimek2015}. One should also take into account unequal losses induced by beam splitter operations and finite lifetime of excitations stored in memory modes. The simplest strategy to mitigate this would be to introduce additional attenuation in order to ensure proper contributions from individual input pulses to the output ports of the receiver. In this case, attainable mutual information calculated in Eqs.~(\ref{Eq:IPPMapprox}) and (\ref{Eq:IHybridOptimal}) would need to be multiplied by the overall power transmission coefficient, which diminishes the superadditivity effect.

\section{Acknowledgements}
We would like to thank J. Nunn for his encouragement to write up this contribution. We acknowledge insightful discussions with U. Andersen, F. E. Becerra, Ch. Marquardt, and M. G. A. Paris. This research was supported in part by the EU 7th Framework Programme projects SIQS (Grant Agreement No. 600645) and PhoQuS@UW (Grant Agreement No. 316244).

\end{document}